\documentstyle[12pt]{article}

\def\hybrid{\topmargin -20pt  \oddsidemargin 0pt
      \headheight 0pt   \headsep 0pt
      \textwidth 6.25in 
      \textheight 9.5in 
      \marginparwidth .875in
      \parskip 5pt plus 1pt   \jot = 1.5ex}

\hybrid

\def\x{\times}

\def\o+{\oplus}
\def\ra{\rightarrow}
\def\beqa{\begin{eqnarray}}
\def\eeqa{\end{eqnarray}}

\newcommand{\p}{\partial}
\newcommand{\ti}{\tilde}
\newcommand{\th}{\theta}
\newcommand{\si}{\sigma}
\newcommand{\ep}{\epsilon}
\newcommand{\al}{\alpha ^{\prime}}
\newcommand{\ba}{\bar}
\newcommand{\resetcounter}{\setcounter{equation}{0}}     

\begin{document}
\thispagestyle{empty}
\rightline{IASSNS-HEP-98-59}
\rightline{HUB-EP-98/39}
\rightline{hep-th/9807008}
\vspace{2truecm}
\centerline{\bf \LARGE The Neveu-Schwarz Five-Brane and its Dual Geometries}

\vspace{1.5truecm}
\centerline{\bf 
Bj\"orn Andreas$^*$\footnote{andreas@physik.hu-berlin.de}\ , \ 
Gottfried Curio$^\dagger$\footnote{curio@ias.edu, supported by
NSF grant DMS9627351}\ and \
Dieter L\"ust$^*$\footnote{luest@physik.hu-berlin.de}}

\vspace{.6truecm}
{\em 
\centerline{$^*$Humboldt-Universit\"at, Institut f\"ur Physik,
D-10115 Berlin, Germany}}

\vspace{.4truecm}
{\em
\centerline{$^\dagger$School of Natural Sciences, Institute for Advanced Study,
 Princeton, NJ 08540}}

\vspace{1.0truecm}
\begin{abstract}
In this paper we discuss two aspects of duality transformations on
the Neveu-Schwarz (NS) 5-brane solutions in type II and heterotic string
theories. 
First we demonstrate that the non-extremal 
NS 5-brane background is U-dual to its
CGHS limit, a two-dimensional black hole times $S^3\times T^5$;
an intermediate step is provided by the 
near horizon geometry which is given by the three-dimensional
$BTZ_3$ black hole 
(being closely related to $AdS_3$) times $S^3\times T^4$.
In the second part of the paper we discuss the T-duality between  
$k$ NS 5-branes and the Taub-NUT spaces respectively ALE spaces, which
are related to the resolution of the $A_{k-1}$ singularities
of the  non-compact orbifold ${\bf C}^2/{\bf Z}_{k}$.
In particular 
in the framework of $N=1$ 
supersymmetric gauge theories related to brane box constructions 
we give the metric dual to two sets of intersecting NS 5-branes. In this way
we get a picture of a dual orbifold background
${\bf C}^3/ \Gamma$
which is fibered together
out of two $N=2$ models ($\Gamma={\bf Z}_k\times {\bf Z}_{k'}$). 
Finally we 
also discuss the intersection of NS 5-branes with D branes, which can serve
as probes of the dual background spaces.
\end{abstract}

\bigskip \bigskip
\newpage

\section{Introduction}

The NS 5-brane [\ref{fivebr}] is one of
the  first string soliton solutions,
which can be constructed 
both for the type IIA/B superstrings as well as for
the heterotic string.  In certain limits there
exists a CFT description of the NS 5-brane (see also [\ref{rey}]). 
In type IIB the NS 5-brane is S-dual to the D 5-brane [\ref{schwarz}].

The NS 5-brane plays an important 
role in the construction of gauge theories from branes [\ref{hw}].
In this context
it serves 
as a kind of background for the D-branes on which the gauge fields
live, since the NS 5-branes are heavy and the D-branes are light.
It is known for some time that $k+1$ parallel 
NS 5-branes are T-dual to the ALE space with
$A_k$ singularity
[\ref{OV}], which is the local geometry of a K3 around a singularity.
So now ALE is the background which is probed by D branes, which are then
also called fractional branes [\ref{frac}].
If one considers intersecting NS 5-branes one gets a background
for N=1 models (brane boxes) [\ref{boxes}]. After T-duality one gets a space
with a local  ${\bf C}^3/\Gamma$, $\Gamma={\bf Z}_k\times {\bf Z}_{k'}$ 
singularity. This describes the situation
of a specific Calabi-Yau manifold around a singularity. 
The T-duality between the Hanany-Witten set up and the fractional branes
was already recently discussed for $N=2$ 
space-time supersymmetry in [\ref{KLS}] and for the $N=1$ brane box models
in [\ref{HU}].
We will discuss several aspects of the duality among NS 5-branes and ALE.

Another aspect arises from the observation that D or M brane solutions
are in fact U-dual to their own horizon geometry [\ref{Boon},\ref{SS}]. 
This provides a relation to supergravity on anti-de Sitter spaces 
and, in the gauge theory picture,
to a corresponding large N limit [\ref{M}]. Via S-duality one
expects an analogous behaviour for NS 5-branes.
In fact we will explicitly demonstrate that the NS 5-brane 
background is U-dual
to its CGHS limit via the intermediate step of $BTZ_3$
[\ref{BHTZ}].

This paper is organized as follows. 
In the next section we present a short summary on the asymptotic
geometry change of string background spaces using the heterotic [\ref{sdual}]
or  respectively the type IIB 
[\ref{schwarz}] S-duality group. Then in
section 3, applying the previous
discussion,  we show
the equivalence of the non-extremal NS 5-brane string background to its
own horizon geometry, namely the
BTZ black hole respectively the CGHS limit [\ref{CGHS}], via U-duality.
In section 4 we will turn the discussion to the T-duality
between  NS 5-branes and the ALE type of spaces.
Here we especially focus on the 
construction of the dual background spaces for  intersecting NS 5-branes
versus 
six-dimensional non-compact orbifolds
${\bf C}^3/\Gamma$
which describe the local neighborhood
around the singularities of a certain Calabi-Yau 3-fold.

\section{Asymptotic Geometry Change}

\resetcounter

Recently it has been observed [\ref{Boon},\ref{SS}] that 
via a sequence of duality
transformations the metrics of M 2-, D 3- and M 5-branes with
flat asymptotic geometry can be transformed to their own horizon
geometries which correspond to the asymptotic non flat spaces $AdS_4\times
S^7$, $AdS_5\times S^5$ and $AdS_7\times S^4$, respectively.
Here, one element within this series of duality transformations is
a certain change, $S$, of coordinates  $(u,v)$ 
which has the form an $Sl(2,{\bf R})$
transformation:
\begin{equation}
S:\qquad \pmatrix{v\cr u\cr}\rightarrow 
\pmatrix{1&-h\cr 0&1\cr}\pmatrix{v\cr u}.
\end{equation}
This transformation has the effect that it shifts to zero the constant
part in the harmonic functions of the above p-brane metrics.

Another way to perform the described geometry change 
by removing the constant part in the metric goes back to
the already older work of [\ref{B}], where the 
asymptotically flat four-dimensional Taub-NUT
space of the KK monopole was transformed into the non-flat 
ALE manifold via a
$TST$ duality transformation. In this context, $S$ now means a genuine
strong-weak coupling duality plus an 
axionic shift transformation 
which is again an element of $Sl(2,{\bf R})$. This approach was recently
applied  to D branes [\ref{BB}], in particular by dualizing 
D 3-branes to intermediates D instantons, and also to heterotic black holes
in four and two dimensions [\ref{camo}].

In the this chapter we like to apply analogous techniques to
transform the metric of the NS 5-brane to its own horizon geometry via
eliminating the constant part in the metric. For the extremal
NS 5-brane this brings us to the socalled throat limit which can
be described by an WZW type superconformal field theory. For the
non-extremal NS 5-brane the sequence of duality transformations will
bring us to the socalled CGHS limit [\ref{CGHS}]. 

First let us recall some facts about the strong-weak coupling S-duality
group $SL(2,{\bf R})$ can be used to perform asymptotic geometry changes.
The discussion goes in parallel both for the heterotic as well as
type II NS 5-brane.


The two well-known frameworks for the non-perturbative group of S-duality
symmetries are the 4-dimensional heterotic string on $T^6$ and the 
10-dimensional  type IIB
string. The group $Sl(2)$  operates 
fractionally linear
on the combination of axion and coupling constant $e^{\phi}$,
where in the heterotic case we have
$S=a+ie^{-2\phi}$ 
with $a$ being the Hodge dual of the four-dimensional $B_{\mu\nu}$-field;
in the type IIB case $S$ is given by 
$S=l+ie^{-\phi}$
with $l$ being the RR scalar field.

We will see that for metrics flat in the Einstein frame the asymptotic
geometry change (in the string frame) is caused, in the end, by the
possible axion shift (in the hetrotic string) resp. RR scalar shift
in the 7-brane picture (for the type IIB string), i.e. the well-known
(F-theory) monodromy in the transversal complex plane (around the 
``singular elliptic fibre'' at the 7-brane position, cf. the stringy cosmic
string [\ref{GSVY}], and [\ref{V}]).

The $Sl(2)$ S-duality group has the 
generators
$\omega_1:=
{\left( \begin{array}{cc} 1&1\\0&1 \end{array}\right)}$,
$\omega_2:={\left( \begin{array}{cc} 0&1\\-1&0 \end{array}\right)}$.
The interpretation of these elements with respect to $S=x+iy$ 
is well-known:\footnote{In our actual computations we are going to the
Euclidean space where the complex $S$-field is replaced by $S_{\pm}=x\pm y$.}
$\omega_1$ gives the shift in $x$ (axion for the heterotic string resp. 
RR-scalar for type IIB), $\omega_2$ gives the coupling constant inversion
$y\ra -\frac{1}{y}$ (for vanishing $x$). One obvious question arises:
besides the inversion element $\omega_2$ and the upper triangular elements
$\omega_1(B)$ one has the lower triangular elements which are conjugates 
of the upper triangular elements by the non-perturbative inversion element
$\omega_2$ 
(we write $\omega_1(B)=
{\tiny \left( \begin{array}{cc} 1&B\\0&1 \end{array}\right)}$):
${\left( \begin{array}{cc} 1&0\\C&1 \end{array}\right)}=
\omega_2\omega_1(-C)\omega_2^{-1}$.

In the following we will elucidate their connection with the notion of
{\em asymtotic geometry change} in both of the mentioned string theories,
the heterotic string and the type IIB.

A suitably normalised element of the relevant subgroup is the shift 
\beqa
S_{\rm shift}={\left( \begin{array}{cc} 1&0\\-1/2&1 \end{array}\right)}
\eeqa
For reasons explained in section (4.1)  we will be especially interested
in the case where $x=y$. 
In that case $S_{shift}$ has the effect of
mapping again to a $S'$ of $x'=y'$
where the effect on $y=e^{-j\phi}$  
($j=2,1$ for the heterotic resp. IIB 
case) can be described as
\beqa
S_{shift}: y^{-1}\ra y'^{-1}=y^{-1}-1
\eeqa
(and the same for $x$) which reflects the nature of the involved
(upper triangular=ordinary) shift element conjugated by
the inversion element.
 
In order to eventually perform the desired geometry change the $S$-duality
transformation has to be combined with further elements of the U-duality
group. For the heterotic string the additional elements are just T-duality
transformations (see section (4.1) for
the concrete treatment of T-duality), such that one ends up with the combined
transformation of the form $TS_{\rm shift}T$. 
On the other hand, for the type II
geometry a more involved U-duality transformation is
necessary, namely one has to consider the sequence 
${\cal T}^{\prime}S_{\rm shift}{\cal T}^{\prime}$ with
${\cal T}^{\prime}:=T_iS_{\rm inv}T_{WV}$.
Here $T_i$ is a T-duality transformation along the direction i within
the world volume of the 5-brane  and $T_{WV}$ denotes the T-dualisation
of all world volume directions. 
The reason why one has to use a more involved sequence ${\cal T}^{\prime}$  
in the type II string compared to the simple T-duality in the heterotic case
relies 
in the fact that for type II strings one is transforming the system to 
intermediate D instantons.
Then 
one has a space-time interpretation for
$S_{shift}$, acting like on the heterotic axion now on the 
type IIB RR-scalar field
respectively,
which amounts to an {\em asymtotic geometry change} by
`deleting the ``1" in the harmonic function' (for the 4D heterotic
space-time geometry resp. the 4D transversal directions to the 5-brane)
(cf. [\ref{BB}]). 
Namely the background to which $S_{shift}$ is applied
to achieve this effect has to be {\bf flat in the Einstein frame}, so
that one has in the string frame
\beqa
ds^2=V(\xi)d\xi^2\nonumber\\
e^{j\phi}=V(\xi)
\eeqa
As $y^{-1}=V$ the desired effect in the metric follows.\\ \\



%

Let us describe briefly the process of asymtotic geometry
change from the extremal type II NS 5-brane to its horizon
geometry, i.e. to the socalled throat limit. As said before, the
throat limit corresponds to
`deleting the ``1" in the harmonic function' $H_5$
(this time in the transversal geometry of the 5-brane) and is achieved by
the element of the U-duality group 
${\cal T}^{\prime}S_{shift}{\cal T}^{\prime}$ where
${\cal T}^{\prime}:=TS_{inv}T_{WV}$ which changes the type IIA NS 5-brane
to a type IIB D(-1) brane. Note that the (-1) brane carries 
the electric charge (measured by an integral over the $S^9$ of its 
transversal space, i.e. all of space-time) for the RR-scalar, whereas
the Hodge-dual 7-brane carries the magnetic charge (measured by the 
$l_{RR}$ upper triangular monodromy in ${\bf C}_{transversal}$).

In the sequence ${\cal T}^{\prime}$ the NS-NS B field is
by S-duality of type IIB mapped to a RR B field the Hodge-dual 6-form
of which becomes after $T_{WV}$ the RR scalar. 
As the 5-brane metric
is {\em flat in the Einstein frame}, which means essentially that
$e^{2\phi}=H_5$, 
the shift in the inverse of $y=e^{-\phi}=H_5^{-1}$ 
leads to the desired
effect in the metric (which in the string frame is just given by the
harmonic function times the flat metric).
So the metric is indeed dually mapped to some `subsector' of itself, the 
near-horizon geometry. 
So we see that, if we include the
conjugating inversion element $\omega_2$ of $S_{shift}$ in the conjugation
process ${\cal T}^{\prime}$, that it is in the end really the (upper 
triangular) monodromy of the RR-scalar in the transversal complex plane
of the dual 7-brane which causes the asymptotic geometry change. 
The 7-brane is magnetically charged for the RR-scalar 
(which is ${\cal T}^{\prime}$-dual to the magnetic NSNS B-field 
of the NS 5-brane in type IIA we started with), which is detected
by the mentioned monodromy related to the stringy cosmoc string resp.
the singular elliptic fibre of the associated F-theory situation.\\ \\


\section{U-duality of the NS five-brane with its CGHS limit}

Now we want to describe in more detail the geometry change from the
non-extremal type NS 5-brane to the socalled CGHS limit [\ref{CGHS}].
For this below a two step process is described which shows 
that the CGHS-limit,\footnote{This limit is relevant for the new 
QFT's in D=6 and D=5 [\ref{S}].}
$BH_2\x S^3$, (which can be interpreted as an $\al$ exact solution) 
of the near extremal NS 5-brane (cf. for example [\ref{MS}]),
\beqa
NS_5\longrightarrow BH_2\x S^3_{{\cal Q}_5},
\eeqa
is actually U-dual to it via the combination of the U-duality [\ref{SS}]
\beqa
NS_5\simeq BTZ_3\x S^3 \x T^4
\eeqa
with the T-duality [\ref{HW}]
\beqa
BTZ_3\x S^3 \x T^4\simeq BH_2\x S^3\x (S^1\x T^4).
\eeqa
As this comes down to `deleting the additive constant $``1"$ in the
harmonic function', i.e. a {\em change of asymptotic geometry}, 
this compares nicely with a corresponding duality 
(TST, the classical Ehlers-Geroch transform) between two purely
gravitational backgrounds [\ref{B}], which one also gets 
by `deleting the additive constant $``1"$ in the
harmonic function', namely between the ALE instanton and the multi 
Taub-NUT space. This is a transformation operating purely in the
4-dimensional transversal space whereas the dualities
shown above make use of transformations in the world-volume sector, too. 
Nevertheless the comparison matches nicely as the 5-brane is well known
to be T-dual to the ALE-space [\ref{OV}] - but on a transversal direction
of the 5-brane compactified on a circle (corresponding with the 
$S^1_{\tau}$-fibration of the ALE-space), which on closer inspection 
has some quite non-trivial subtleties [\ref{GHM}]. These issues are 
described in a section 3.

\subsection{The CGHS limit NS 5 $\ra BH_2 \x S^3$}

For the near-extremal  
NS 5-brane in type IIA with its world-volume 
compactified on $S^1_1\x T^4_{2345}$
(we suppress the flat spatial 
world-volume directions) one has [\ref{MS}]
\beqa
ds^2&=&-(1-\frac{r_0^2}{r^2})dt^2+(1+\frac{{\cal Q}_5\al}{r^2})
(\frac{dr^2}{1-\frac{r_0^2}{r^2}}+r^2d\Omega_3^2),\nonumber\\
e^{2\phi}&=&e^{2\phi_{\infty}}(1+\frac{{\cal Q}_5\alpha ^{\prime}}{r^2}),
\eeqa
with 
$H={\cal Q}_5\epsilon _3$ 
(cf. sect. (3.2)). 
Let us introduce the near-horizon coordinate $\si$ with
\beqa
r=r_0\cosh \si ,
\eeqa
the non-extremality-parameter $\alpha_5$
\beqa
\sqrt{{\cal Q}_5}=\frac{r_0}{\sqrt{\al}}\sinh \alpha_5 ,
\eeqa
and the energy density parameter 
\beqa
\mu =\frac{r_0^2}{g^2\al} ,
\eeqa
(here $g:=e^{\phi_{\infty}}$) which occurs [\ref{MS}] in the 
(string-frame)
energy per unit 5-volume\footnote{Here $V_5=(2\pi)^5R_1R_2R_3R_4R_5$.}
$M_5:=\frac{M}{V_5}=
\frac{1}{{\al}^3(2\pi)^5}(\frac{{\cal Q}_5}{g^2}+\mu)=
\frac{1}{{\al}^3(2\pi)^5}\mu \cosh^2 \alpha_5$. 
Then one gets
\beqa
ds^2&=&-\tanh ^2 \si dt^2 +(\mu g^2\cosh^2 \si+{\cal Q}_5)\al 
(d\si^2+d\Omega_3^2) ,\nonumber\\
e^{2\phi}&=&g^2+\frac{{\cal Q}_5}{\mu\cosh^2 \si} .
\eeqa
One sees that
making the $g\ra 0$ limit (CGHS-limit [\ref{CGHS}]), 
while keeping the energy density parameter
$\mu$ at order one,
corresponds to `deleting the 
additive 
constant $``1"$ in the harmonic function' $1+\frac{{\cal Q}_5\al}{r^2}$;
at the same time this causes the decoupling of the $S^3$ sector 
with $ds^2_{S^3}={\cal Q}_5 d\Omega_3^2$ and $H={\cal Q}_5\epsilon_3$, 
leading to the 2-dimensional black hole times the $SU(2)$ WZW model:
\beqa
ds^2&=&-\tanh ^2 \si dt^2 +{\cal Q}_5\al d\si^2,\nonumber\\
e^{2\phi}&=&\frac{{\cal Q}_5}{\mu\cosh^2 \si}.
\eeqa
\\ \\
For the sake of later comparison let us transform the coordinates back by
$\bar{r}:=\sqrt{\frac{\mu}{{\cal Q}_5}}\cosh \sigma=\bar{r_0}\cosh \sigma$:
\beqa
ds^2&=&-(1-\frac{\bar{r}_0^2}{\bar{r}^2})dt^2+
\frac{{\cal Q}_5\al}{\bar{r}^2}
\frac{d\bar{r}^2}{1-\frac{\bar{r}_0^2}{\bar{r}^2}},\nonumber\\
e^{2\phi}&=&\frac{1}{\bar{r}^2} .
\eeqa

Let us remark that $g\ra 0$ implied here that $r_0\ra 0$ as $\al$ is kept
fixed here. [\ref{M}] makes $\al \ra 0$ instead of $g\ra 0$, 
apart from that 
similar reduction to the near-horizon region.\\ \\

\subsection{The duality NS 5 $\simeq BTZ_3\x S^3\x T^4$}

Let us start again with the NS 5-brane in type IIA with its world-volume 
compactified on $S^1_1\x T^4_{2345}$ in the notation
of [\ref{SS}] (set $H_1=1$ for now; let $dy^2:=dx_2^2+\cdots +dx_5^2$):
\beqa
ds^2&=&\frac{1}{H_1}[-(1-\frac{r_0^2}{r^2})dt^2+dx_1^2]+dy^2+
(1+\frac{{\cal Q}_5 \al}{r^2})(\frac{dr^2}{1-\frac{r_0^2}{r^2}}+
r^2d\Omega_3^2),\nonumber\\
e^{2\phi}&=&\frac{1}{H_1}(1+\frac{{\cal Q}_5 \al}{r^2}),\nonumber\\
H_{ijk}&=&\frac{1}{2}\ep _{ijkl}\p _l
(1+\coth \alpha_5 \frac{{\cal Q}_5 \al}{r^2}) .
\eeqa
Note that $\phi$ is shifted by $\phi_\infty$ compared to
the previous equations.

So we are in the case\footnote{Also $\alpha_K={\cal Q}_K=0, H_K=1$.}
${\cal Q}_1=\alpha_1=0$ 
of [\ref{SS}] where  $H_i=1+\frac{{\cal Q}_i \al}{r^2}$ and
${\cal Q}_i=\frac{r_0^2}{\al}\sinh ^2 \alpha _i$ with $i=1,5$. 
Note that because of the 
$\coth \alpha_5=\sqrt{1+\frac{r_0^2}{{\cal Q}_5\al}}$
in $H$ only for
the extremal case of $\alpha_5$ very large 
one has an axionic instanton. (Note that $g\ra 0$ implies 
(by the $\mu$ condition) $r_0\ra 0$ and so $\alpha_5\ra \infty$.)\\ 

This is U-dual 
via ${\cal T}^{\prime}S_{\rm shift}{\cal T}^{\prime}$ 
(with ${\cal T}^{\prime}:=
T_1ST_{1234}ST_5$, 
cf. [\ref{SS}])\footnote{We use here as a technical device the 
shift interpretation as
coordinate change in the fundamental wave (reached from the IIB D-string,
which we got from the IIA NS 5 brane
after the part $T_1ST_{1234}$ of ${\cal T}$, by the
remaninig part $ST_1$); this is a technical alternative to the 
interpretation of the the wave in a 12D sense 
(the D-(-1) brane [\ref{T}]) which one gets after part $T_{05}$ 
which follows in the construction of ${\cal T}^{\prime}=TST_{123405}$ 
described in the introduction. This gives the shift (deleting the ``1'')
for $H_5$; it has actually to be coupled with a similar procedure for
$H_1$ (cf. [\ref{SS}]).}
to a configuration with $H_1=\frac{r_0^2}{r^2}=H_5$
\beqa
ds^2=\frac{r^2}{r_0^2}[-(1-\frac{r_0^2}{r^2})dt^2+dx_1^2]+dy^2+
\frac{r_0^2}{r^2}(\frac{dr^2}{1-\frac{r_0^2}{r^2}}+r^2d\Omega_3^2)
\eeqa
and $e^{2\phi}=1, B_{01}=\frac{r^2}{r_0^2}-1, 
H_{ijk}=\frac{1}{2}\ep _{ijkl}\p _l(\frac{r_0^2}{r^2}-1)$.

Note that the ${\cal Q}_5$-dependence, which seems to be lost, 
is still kept as the rescaling $R_5\ra R_5 \cosh \alpha_5$ has happened.
Because of the $R_5$-rescaling 
the 3-dimensional Newton constant
is not\footnote{Here $V_{T^4}=(2\pi)^4R_2R_3R_4R_5$ is the volume in the
beginning.} 
${\cal G}_N^{(3)}=\frac{G_N^{(10)}}{V_{T^4}\cdot (r_0^3\Omega_3)}$
but $G_N^{(3)}=\frac{{\cal G}_N^{(3)}}{\cosh \alpha_5}$; so
$G_N^{(3)}$ is here a function of $r_0$ and ${\cal Q}_5$.

Now by effectively
`deleting the additive constant $``1"$ in the
harmonic function' $H_5$ 
(besides changing ${\cal Q}_5\al$ to $r_0^2$)
the $S^3$ sector has decoupled where one has an $S^3$ of radius $r_0$
with $ds^2_{S^3}=r_0^2d\Omega_3^2$ and $H=r_0^2\ep_3$.
This leads to the structure $BTZ_3\x S^3\x T^4_y$ with the metric 
(besides $e^{2\phi}=1, B_{t\varphi}=r_0(\frac{r^2}{r_0^2}-1)$; 
$\varphi:=x_1/r_0$)
\beqa 
ds^2_{BTZ}=-(\frac{r^2}{r_0^2}-1)dt^2+r^2d\varphi^2+
\frac{dr^2}{\frac{r^2}{r_0^2}-1}.
\eeqa

After the rescaling
$t \ra ct,\; \varphi \ra c\varphi ,\;
r \ra c^{-1}r,\; 
r_0 \ra c^{-1}r_0=\sqrt{{\cal Q}_5\al}$
of the metric by $c=\frac{r_0}{\sqrt{{\cal Q}_5\al}}$
it takes with $M_3=c^2=r_0^2/({\cal Q}_5\al)$ the form
\beqa
ds^2=-M_3(\frac{r^2}{r_0^2}-1)dt^2+r^2d\varphi ^2+
\frac{1}{M_3}\frac{dr^2}{\frac{r^2}{r_0^2}-1} .
\eeqa

\subsection{The non-compact untwisting $BTZ_3\simeq BH_2\x S^1$}

To describe the 3-dimensional BTZ black hole [\ref{BHTZ}] in its relation to 
anti-de Sitter space note first that $AdS_3$ is 
\beqa
-x_0^2-x_1^2+x_2^2+x_3^2=-l^2
\eeqa
in the flat space of signature $( - - + + )$
\beqa
ds^2=-dx_0^2-dx_1^2+dx_2^2+dx_3^2 .
\eeqa
To get the physical coordinates for the black hole of mass $M_3$ with 
horizon at $r_0$ one makes first the coordinate change to 
$r, \varphi$ and $t$ (with $l^2=\frac{r_0^2}{M_3}$,
$\frac{r^2}{M_3}=x_1^2-x_2^2, 
e^{\varphi \sqrt{M_3}}=\frac{\sqrt{M_3}}{r}(x_1+x_2)$) 
\beqa
x_0&=&\frac{r_0}{\sqrt{M_3}}\sqrt{1-\frac{r^2}{r_0^2}}
\cosh (t \frac{M_3}{r_0}),\nonumber\\
x_1&=&\frac{r}{\sqrt{M_3}}\cosh \varphi \sqrt{M_3},\nonumber\\
x_2&=&\frac{r}{\sqrt{M_3}}\sinh \varphi \sqrt{M_3},\nonumber\\
x_3&=&\frac{r_0}{\sqrt{M_3}}\sqrt{1-\frac{r^2}{r_0^2}}
\sinh (t \frac{M_3}{r_0}),
\eeqa
and then identifies $\varphi$ with period $2\pi$ 
to get $BTZ_3$ from $AdS_3$ [\ref{BHTZ}],[\ref{HW}]
leading to the metric 
(besides $e^{2\phi}=1, B_{\varphi t}=\sqrt{M_3}r^2/r_0$) 
\beqa
ds^2=-M_3(\frac{r^2}{r_0^2}-1)dt^2+\frac{1}{M_3}\frac{dr^2}
{\frac{r^2}{r_0^2}-1}+r^2d\varphi ^2 .
\eeqa
 
If one makes a T-duality along the $\varphi$-direction, where one has a 
translational symmetry, one gets (after the further 
coordinate change\footnote{As $\varphi$ is periodic, so are $\ti{t}$ and 
$\ti{\varphi}$; one actually works then on the covering space, to avoid 
CTC's.}
$\ti{t}=\varphi/r_0, \ti{\varphi}=\sqrt{M_3}t+\ti{t}$) [\ref{HW}]
\beqa
\ti{ds}^2&=&-(1-\frac{r_0^2}{r^2})d\ti{t}^2+
\frac{r_0^2/M_3}{r^2}\frac{dr^2}{1-\frac{r_0^2}{r^2}}+
d\ti{\varphi}^2,\nonumber\\
e^{2\phi}&=&\frac{1}{r^2},
\eeqa
with $B_{\ti{\varphi}\ti{t}}=0$.
This is the 2-dimensional black hole times $S^1$.
As $AdS_3$ is\footnote{The $\det g=1$ condition for 
$g={\tiny \left( \begin{array}{cc} a&b\\c&d \end{array}\right)}
\in Sl_2({\bf R})$ with 
$x_{0/3}=\frac{b\pm c}{2}, x_{2/1}=\frac{a\mp d}{2}$ translates to
$l=i$ causing the signature change.}
(up to signature) $SL(2,{\bf R})$, and the 2-dimensional black hole
is the $SL(2,{\bf R})$ WZW model with a $U(1)$ gauged [\ref{wbh}], we see that
the T-duality above is just a non-compact version of the 
T-duality-`untwisting' of $S^3=SU(2)$ to $S^2\x S^1=SU(2)/U(1)\x U(1)$. 

As $r_0^2/M_3={\cal Q}_5\al$ we find coincidence 
with the crucial prefactor
of the $d\bar{r}^2$ term in the CGHS limit.

\section{Some aspects of the 
T-duality of the Taub-NUT spaces with the five-brane}

\resetcounter

We consider the T-duality for the ALE spaces with the five-brane.
This was first made plausible by the argument of [\ref{OV}] that under
a fibrewise T-duality for an elliptically fibered $K3$ with $A_{k-1}$
singularity the monodromy for the complex structure parameter of the
elliptic fibre (caused by the singularity) goes over to the monodromy 
for the K\"ahler parameter which gives the crucial H charge $k$. On closer
inspection [\ref{GHM}] this has some non-trivial points described below.
(For a treatment from an other perspective cf. [\ref{Sen}].)\footnote{
[\ref{Sen}] argues that at a point in moduli space, where the 
center positions of the Taub-NUT metric (=KK monopole of IIa resp. M)
merge (the critical point for gauge symmetry enhancement), and near
the singularity (where the membranes wrapping the vanishing $S^2$ become 
massless giving the new non-abalian gauge bosons) the 
pol terms dominate so one can effectively neglect the ``1" in the 
harmonic function leading to the ALE situation; then a TST transformation
is made to the system of coalescing D6 branes in IIA 
(=KK monopole of M=Taub-NUT) where the mentioned membranes become the
stretched strings between the D6 branes.}
The T-duality between ALE spaces and axionic instantons was also discussed
in [\ref{bianchi}].

\subsection{Case of one isometric direction}

We will consider the case where we perform the T-duality with respect
to one isometric direction.
The spaces we are going
to start with are the purely gravitational backgrounds given by the ALE
and Taub-NUT spaces.
Of these the ALE spaces describing the resolutions 
of the $A_{k-1}$ singularities
are complex two-dimensional non-compact relatives of $K3$, i.e. 
non-compact Ricci-flat hyperkaehler manifolds. The ALE manifold of the 
$A_{k-1}$ series corresponds to the metric given by the Gibbons-Hawking 
multi-center ansatz
\beqa
ds^2=V(\vec{x})d\vec{x}^2+
V^{-1}(\vec{x})(d\tau+\vec{\omega}\cdot d\vec{x})^2
\eeqa 
with the self-duality condition 
$\vec{\nabla}V=\vec{\nabla}\x \vec{\omega}$, where we are in the case 
$\ep =0$ of
\beqa
V=\ep +\sum_{i=1}^{k}\frac{1}{|\vec{x}-\vec{x_i}|}  
\eeqa
This space $M_{k-1}$ is the smooth resolution of the singular variety
$xy=z^k$ in ${\bf C}^3$ of type $A_{k-1}$ 
with $\partial M_{k-1}=S^3/{\bf Z}_k$.
The singular situation corresponds to the pol-terms coalescing:
$V=\frac{k}{|\vec{x}|}$.
The case $\ep=1$ corresponds to the Taub-NUT spaces.

Now T-duality with respect to the $U(1)$-isometry generated by the Killing
vector $\partial /\partial \tau$ gives with the well-known 
Buscher formula
the conformal flat metric of the extremal NS 5-brane (cfr. eq.(3.8) with
$r_0=0$)
\beqa
ds^2&=&V(\vec{y})(d\tau^2+d\vec{y}^2),\nonumber\\
B_{0i}&=&\omega _i,\nonumber\\
e^{2\phi}&=&V(\vec{y}),
\eeqa
where the self-duality condition for the original metric is now, in the 
new axion-dilaton sector, assuring the condition for an axionic instanton
\beqa
H_{\mu\nu\rho}=
\sqrt{g}{\epsilon _{\mu\nu\rho}}^{\sigma}\partial _{\sigma}\phi .
\eeqa
The H charge is from 
$H_{\mu\nu\rho}=
\sqrt{g}{\epsilon _{\mu\nu\rho}}^{\sigma}\partial _{\sigma}\phi$
with $e^{2\phi}=V=\frac{k}{|\vec{x}|}$ easily seen to be 
$\frac{1}{2\pi^2}\int_{S^3}H=k$.
This shows the appearence of the required $H$ charge $k$.

The heterotic\footnote{A analogous argument can be made
for the type II RR-scalar after dualizing down to the
D-instantons.} axion $a$ 
in the dual geometry is defined by dualizing the dual $B$ field
\beqa
\p a=\pm e^{-2\phi}H_D^*=\frac{1}{2}V^{-2}\p \omega ^*
\eeqa
The axion charge of the T-dual solution is called the nut charge of the 
original gravitational solution. One can say that the S-duality 
group is related to a duality between the electric aspects of the original
gravity background (characterised by the Maxwell field 
$A=V^{-1}(d\tau+\vec{\omega}\cdot \vec{x})$) and the magnetic aspects
(characterised by the nut potential $a$).
Note that the  the isometry we
are using is called `translational' (the main importance of 
this is keeping the SUSY manifest after dualisation), i.e.
the covariant derivative of the Killing vector field is self-dual
which means $(\p S_-)^2=0$. So from $\p S_-=0$ one gets 
$\p V=\p \omega ^*$  
so that $V$ satisfies the 3D Laplace equation
and one has $S_-=0$, i.e. $a=V^{-1}$.
We see that
the Taub-NUT metric with can be mapped 
$S_{\pm}=a\pm e^{-2\phi}=a\pm V^{-1}$
to the ALE space via the shift $V\rightarrow V-1$.\\ \\

Note however
that in $e^{2\phi}=\frac{k}{|\vec{y}|}$ the 3-dimensional harmonic
function $\frac{1}{r_3}$ occurs whereas in the five-brane the 
4-dimensional harmonic function $\frac{1}{r_4^2}$ occurs. The reason is
of course that by doing T-duality in the periodic $\tau$-direction in
Taub-NUT space one arrives at the five-brane with one of its 
four transverse
directions compactified on a circle.
In other words, since
the harmonic function of the original extremal 5-brane
metric also depends on $\tau$, before the T-duality from the
5-brane to the ALE space one
has to {\sl enforce} an isometric
direction by taking the transversal
space to be $R^3_{\vec{x}}\x S^1_{\tau}$ and requiring $H_5$ to be a 3D
harmonic function $H_5=V=1+\frac{{\cal Q}_5}{r_3}$ (here $r_3:=|\vec{x}|$)
in $R^3$ and independent of the $S^1_{\tau}$ direction.
 More precisely if the original 5-brane
metric is scaled so that the $\tau$-direction is very large and the
$\vec{x}$-space looks correspondingly contracted, 
then in the dual space the direction of the corresponding little
circle is `suppressed', 
i.e. the ansatz there leads to a harmonic 
function in only three variables.\\
But note that this is only {\em one} possibility to realise the duality. 
One could also tune directly the merging without making the 
radius $R$ of the 
$S^1_{\tau}$ large; then the dual circle (of radius $\ti{R}=\al/R$) 
is not `suppressed'.\\
More concretely one has in the dual picture the Fourier decomposition 
along the dual circle [\ref{GHM}]
\beqa
e^{2\phi}(\vec{x},\tau)=
\sum_{n\in {\bf Z}} e^{in\tau /\ti{R}}\Psi_n(\vec{x})
\eeqa
where (the $\Psi_n$ are no longer suppressed for $\ti{R}$ 
being no longer small)
\beqa
\Psi_0=e^{2\phi_0}+\frac{k\al}{2r\ti{R}},\;\; 
\Psi_n=\frac{\al}{2r\ti{R}}e^{-|n|r/\ti{R}}e^{-in\tau _0/\ti{R}} \;\;
(n\neq 0)
\eeqa
One can see that the $\Psi_n$ interpolate between the 3-dimensional
and the 4-dimensional harmonic function according to $\ti{R}$ being
very small or very large [\ref{DS}].  
Furthermore the occurence of these momentum modes in the dual picture 
leads to the idea that in the original picture winding modes have to be 
included in the description.\\ 
This can also be understood from the following perspective. In the 
original picture of the ALE space one has besides the 
degree of freedom $\vec{x}_i$ 
(positions of the centers)
also to take into
account the parameters $\int _{\Sigma _i}B$ (the $\Sigma _i$ being the
non-trivial 2-cycles 'between the centers')\footnote{being given by the $
S^1_{\tau}$-fibration over the line in $R^3_{\vec{x}}$ connecting two 
centers; as the $S^1$ shrinks at the centers this is an $S^2$}; 
that they play an important 
role in the game was the insight of [\ref{A}], who showed that actual
gauge symmetry enhancement occurs only, if not only the positions of the
centers merge (giving the $A_{k-1}$ singularity), but also the $B$-field
parameters have the value zero (and not the CFT orbifold value $\pi$; 
this leads to the breakdown of the CFT reasoning, necessary for the
non-perturbative gauge symmetry enhancement). Now each of these 4 real
parameters (for $i=1\cdots, k-1$), the center-distance and the $B$-field
parameter, constitute a hypermultiplet. In the dual picture this 
corresponds to the positions of the 5-branes in the transversal space
$R^3\x S^1$. But the position parameter in the (dual) $S^1$-drection 
breaks the expected isometry in this direction.\\ 
This leads to the 
alternative view on the necessity of including winding modes in the 
original Taub-NUT picture. 
Above we saw this was caused by the actual `occurence'
(being no longer suppressed) of the $S^1$ in the 5-brane picture
which lead to the momentum modes there and so to the
winding modes in the original picture. Here we see that the 
$S^1$-position degree of freedom in the 5-brane picture 
corresponds in the original picture to the $\int _{\Sigma _i}B$
degree of freedom; but one gets the necessary compact
$S^2$-cycles $\Sigma _i$ `between the centers' exactly beacuse the
$S^1$-fibration in the $\tau$ variable in the Taub-NUT space collapses to
circles of zero radius at the center points; and this means that the 
corresponding winding modes there become light and so winding modes
should be included in the description. 

At the end of this section we briefly give the duality transformation
for the non-extremal 5-brane.
Its metric becomes
\beqa
ds^2=-(1-\frac{r_0}{r_3})dt^2+
H_5(\frac{dr_3^2}{1-\frac{r_0}{r_3}}+r_3^2d\Omega_2^2+d\tau^2)
\eeqa
with the H-monopole magnetic field
$H_{\tau\theta \varphi}=-\p_{\theta}B_{\tau \varphi}=
r_3^2\sin \theta \p_{r_3}(1+\coth \alpha_5\frac{{\cal Q}_5}{r_3})$.

Then the non-extremal KK-monopole dual to non-extremal 5-brane is
\beqa
ds^2=-(1-\frac{r_0}{r_3})dt^2+
V^{-1}(d\tau +\vec{\omega}\cdot d\vec{x})^2+
V(\frac{dr_3^2}{1-\frac{r_0}{r_3}}+r_3^2d\Omega_2^2)
\eeqa
with KK magnetic field $F=\p \vec{\omega}$. More precisely
if $\vec{\omega}=A_{\varphi}$ then 
$F_{\theta \varphi}=-\p_{\theta} A_{\varphi}=
r_3^2 \sin \theta\p_{r_3}(1+\coth \alpha_5\frac{{\cal Q}_5}{r_3})$
with ${\cal Q}_5=r_0\sinh^2 \alpha_5$.


In the case of two isometries one will see an effective reduction
by two dimensions down to a function harmonic in two
dimensions, the logarithm.
As there is an effective 2+2 split of the coordinates  and
in view of the hyper-K\"ahler nature of the relevant background, it is
appropriate to describe the situation in an complex superfield formalism,
the general features of which we describe first (cf. for ex. [\ref{KKL}]).

\subsection{Superfield formalism of Buscher duality and two isometries}

The $N=2$ superspace action for one chiral ($\bar{D}_{\pm}U=0$)
superfield
$U$ and one twisted chiral ($\bar{D}_+V=D_-V=0$) superfield $V$
is determined by the real potential function $K(U,\bar{U},V,\bar{V})$
\beqa
S=\frac{1}{2\pi \alpha ^{\prime}}
\int d^2x D_+D_-\bar{D}_+\bar{D}_-K(U,\bar{U},V,\bar{V}) \nonumber
\eeqa
with the target space interpretation ($K_u:=\frac{\p K}{\p u}$)
\beqa 
S_{bos}=-\frac{1}{2\pi \alpha ^{\prime}}
\int d^2x(K_{u\bar{u}}\p ^a u \p _a \ba{u}-
K_{v\bar{v}} \p ^a v \p _a \ba{v}
+\epsilon_{ab}(K_{u\bar{v}}\p ^a u \p ^b \ba{v}+
K_{v\bar{u}}\p ^a v \p ^b \ba{u})) \nonumber
\eeqa
which shows the $G_{\mu\nu}$ and the $B_{\mu\nu}$ part; so, for example,
the $H$ field components become
\beqa
H_{u\ba{u}v}=K_{u\ba{u}v} &,& H_{v\ba{v}u}=K_{v\ba{v}u}\nonumber\\
H_{u\ba{u}\ba{v}}=-K_{u\ba{u}\ba{v}} &,& H_{v\ba{v}\ba{u}}=-
K_{v\ba{v}\ba{u}}.
\eeqa

Furthermore the string equations of motion have to be satisfied (vanishing
$\beta$-function equations). If the central charge deficit (determined 
by the dilaton $\beta$-function) vanishes, one actually has $N=4$ 
supersymmetry in two dimensions.

In general one gets $N=4$ supersymmetry
for a  potential $K$ satisfying the Laplace equation
$K_{u\ba{u}}+K_{v\ba{v}}=0$ (this is the generalization of the 
hyper-K\"ahler
condition for backgrounds including a $B$ field; it is only a sufficient
condition in case of non-trivial dilaton). From the string equations of 
motion one has then
\beqa
\p _u \log K_{v\ba{v}} &=& 2 \p _u \phi ,\nonumber\\
\p _v \log K_{u\ba{u}} &=& 2 \p _v \phi ,
\eeqa
giving $e^{2\phi}\sim K_{u\ba{u}}$ so that the metric is flat in the 
Einstein metric $G^{\mbox{Einst}}_{\mu\nu}=e^{-2\phi}G^{\sigma}_{\mu\nu}$,
i.e. only the axion-dilaton sector is non-trivial and one has
\beqa
H_{u\ba{u}v}=K_{u\ba{u}v}=2e^{2\phi}\p _v \phi
\eeqa
and $d\phi=\pm \frac{1}{2} e^{-2\phi} H^*$, the self-duality condition
for the axion-dilaton sector.

For T-duality one has to assume the existence of (at least) one
$U(1)$-isometry; this cooresponds to a Killing symmetry of the potential 
$K$
\beqa
K=K(u+\ba{u},v,\ba{v}) .\nonumber
\eeqa
The duality will trade in a twisted field $w$ for the untwisted field $u$
by a Legendre transformation leading to the dual potential ($r:=u+\ba{u}$)
[\ref{roc},\ref{KKL}]
\beqa
\ti{K}(r,w+\ba{w},v,\ba{v})=K(u+\ba{u},v,\ba{v})-r(w+\ba{w}) \nonumber
\eeqa
i.e. after the variation w.r.t. $u$: 
$\frac{\delta S}{\delta u}=0\Rightarrow
w+\ba{w}=K_r=K_u$ the independent variables for $\ti{K}$ are 
$w,\ba{w},v,\ba{v}$. As these are now only twisted fields 
(a set containig 
only untwisted fields would of course do it equally well) $\ti{K}$ is a
true K\"ahler potential providing a K\"ahler metric with Ricci-tensor
\beqa
\ti{R}_{i\ba{j}}=-\p _i\p_{\ba{j}}\log \det \ti{G}_{i\ba{j}}=
-\p _i\p_{\ba{j}}\log -\frac{K_{v\ba{v}}}{K_{u\ba{u}}}\nonumber
\eeqa
i.e. the dual background is Ricci-flat for 
$K_{v\ba{v}}\sim K_{u\ba{u}}$.


In the case of 2 translational $U(1)$ Killing symmetries
\beqa
K=K(u+\ba{u},v+\ba{v})\nonumber
\eeqa
the Laplace equation is solved by 
($u:=r_1+i\th,v:=r_2+i\phi, z:=r_1+ir_2$)
\beqa
K(r_1,r_2)=iT(r_1+ir_2)-i\ba{T}(r_1-ir_2)=-2 {\rm Im} T(z)\nonumber
\eeqa
where $T(z)$ is an arbitrary holomorphic function
 and the associated metric is 
\beqa
ds^2= -4{\rm Im}T_{zz}(dud\bar u+dvd\bar v).
\eeqa

For our axionic instanton background consisting of the 5-brane with 
two isometric directions the relevant harmonic function is now just
$H_5=k \log |z|$.

Then the dual metric becomes (with $w=\frac{1}{2}K_u+i\th$) [\ref{KKL}]
\beqa
ds^2&=&\frac{1}{K_{u\ba{u}}}(dw-K_{uv}dv)(d\ba{w}-K_{u\ba{v}}d\ba{v})-
K_{v\ba{v}}dvd\ba{v}\nonumber\\
 &=&{\rm Im} S dzd\ba{z}+\frac{1}{{\rm Im} S}(d\th -S d\phi)(d\th -\ba{S} d\phi)
\eeqa
with $S(z):=-\frac{1}{2} T_{zz}(z)$.

We interpret this as a part, local in the base ($z$-variable), of 
an elliptic fibration (cf. the discussion of the 
stringy cosmic string [\ref{GSVY}]), 
which is thus dual to the axionic instanton 
background we started with. In [\ref{GSVY}] also global issues in the 
base were treated making $S$ a true well-defined modular invariant by 
multyplying it by an $\eta$-function term, i.e. $S(z)$ is just a local 
version of $\tau (z)$. If one specialises to an $A_{k-1}$ singularity
one has $k$ cosmic strings at $z=0$, i.e. $j(\tau (z))=\frac{1}{z^k}$. 
Now at $\tau \approx i\infty , j\approx \infty$ one has 
$j(\tau )\sim e^{-2\pi \tau (z)}$ or $\tau (z)=-\frac{k}{2\pi }i \log z$, 
so
\beqa
{\rm Im} S(z)=-\frac{k}{2\pi } \log |z| .
\eeqa

Note that the $H$ charge of the original axionic instanton background
is from $H_{u\ba{u}v}=K_{u\ba{u}v}=2e^{2\phi}\p _v \phi$ found to be 
proportional to $n$ as the dilaton was
$e^{2\phi} \sim K_{u\ba{u}}=-2 {\rm Im}  T_{zz}=4{\rm Im S}$, i.e. 
\beqa
e^{2\phi}\sim k \log |z|
\eeqa
which shows the consistency of the interpretation.

Note that the ALE description (which is local around the (resolved) 
singular point) is related to the resolution of the singularity, 
whereas the description given here in the fibration picture (which is
local in the base around the (desingularised) fibre, but global in the 
fibre) is related to the deformation of the singularity to an elliptic
fibration of smooth total space. 

\subsection{Intersection of two NS 5-branes -- $N=1$ brane boxes}

In the gauge-theory-from-branes setup the NS 5-branes are considered 
to be heavy relative to the D4-branes whose world-volume gives the 
gauge theory. So the NS 5-branes (later, after T-duality, 
the KK-monopoles, resp., if one is interested in the singular 
situation  at the 
neighborhood of the singularity, the ALE spaces in the 
transversal dimensions) constitute 
the `background', the D4-branes (later, after T-duality,
the fractional D3-branes) 
are the `probes'.  

So there are really two levels of consideration here:
first the gauge  theory  where gravity is turned off and the 
light D3-branes; second there is  
a background, probed by the
D3-branes, of ALE/KK-monopoles, the T-dual of the
NS 5-branes which are considered to be heavy. In the case of
$N=2$ space-time supersymmetry with
parallel NS 5-branes the background is just the well-known 
ALE/KK-monopole space; in the $N=1$ case it is a background of two
ALE/KK-monopole spaces fibered together over a common $R^2$ direction.
This back ground arises as the T-dual of intersecting NS and NS'
5-branes which build the socalled brane boxes of [\ref{boxes}],
as we will discuss in the following.

If on the other hand the backreaction of
the D3-branes on the background is included, the former NS 5-branes
(resp. their T-duals) become dynamical and it is appropriate to give a 
common metric for the total brane system. This will be the
topic of the next subsection.


Let us describe the metric for the $N=1$ situation of a 
${\bf C}^3/\Gamma$ singularity, $\Gamma={\bf Z}_k\times
{\bf Z}_{k^{\prime}}$, (probed by the D3-branes)
[\ref{HU}] in the case of ``adding up" two
ALE spaces. Let us forget about 
the D-branes and just concentrate on the two now non-parallel 
NS 5-branes. Specifically to compute the metric
of the NS-NS' 5-brane system one starts  one step earlier with 
k D5-branes in 012345 and $k'$ D5-branes in 012367
with compact directions
4 and 6. This has the metric [\ref{BBJ}] 
(with $e^{-2\phi}=H_{5}H_{5^{\prime}}$)
\beqa
ds^2=\frac{1}{\sqrt{H_{5}H_{5^{\prime}}}}ds_{0123}^2+
\sqrt{\frac{H_{5^{\prime}}}{H_{5}}}ds_{45}^2+
\sqrt{\frac{H_{5}}{H_{5^{\prime}}}}ds_{67}^2+
\sqrt{H_{5}H_{5^{\prime}}}ds_{89}^2
\eeqa
Then this is S-dualised to k NS 5-branes in 012345 and 
$k'$ NS' 5-branes in 012367 giving 
\beqa
ds^2=ds^2_{0123}+H_{5^{\prime}}ds_{45}^2+H_5ds_{67}^2+
H_{5}H_{5^{\prime}}ds_{89}^2
\eeqa
with $e^{2\phi}=H_{5}H_{5^{\prime}}$.

Then one makes T-dualities in the compact directions  4 and 6 giving
(at the singularity) an $A_{k-1}$ in 6789 and an $A_{k'-1}$ in 4589.\\
This ``adding up'' of the two ALE spaces is difficult to perform in the 
usual representationfor the ALE metric which has a 3+1 split in the 
coordinates. Instead one would like to have a representation which 
isolates the singularity in a $2+2_{89}$ description. This is provided
by the stringy cosmic string description of section
(4.2) which gives the $A_{k-1}$ 
singularity in an elliptic fibration over ${\bf C}=R_{89}$ (so here
the directions 5 and 7 are considered to be compact too; the respective
``$\tau$-circles" of the ALE spaces 
$A_{k-1}$ in 6789 and an $A_{k'-1}$ in 4589
are well known to shrink to zero radius at the, in the merging limit
common, center; they are the vanishing $S^1$'s in the elliptic 
fibration). 

So one assumes that $H_5$ (and correspondingly for $H_{5^{\prime}}$), 
which - to make $T_6$ - was assumed to
be independent of $x_6$ and just living as a harmonic function in 789,
is now actually independent also of $x_7$ and so lives just as a 
logarithmic function in 89 (cf. eqn. (4.16)  above:
$e^{2\phi}\sim k \log |z_{89}|$).\\

Because of the 
fibration structure now this type of representation of the ALE metric
is - in contrast to the 3+1 representation - easily ``added up". So this 
extends the $N=2$ supersymmetric
case with the (local in the base) description of an 
$A_{k-1}$ singularity
of an elliptic fibration (over ${\bf C}_{89}$) to the $N=1$ 
supersymmetric case of a 
description\footnote{G. C. thanks A. Uranga for discussion on this point}
of the singularities of the doubly elliptically fibered
Calabi-Yau space\footnote{$dP_9=
\frac{1}{2}K3$ is the elliptically fibered surface $dP_9={\tiny 
\left[\begin{array}{c|c}P^2&3\\P^1&1\end{array}\right]}$.}
$CY^{19,19}={\tiny 
\left[\begin{array}{c|cc}
P^2_{z_{45}}&3&0\\P^1_{z_{89}}&1&1\\P^2_{z_{67}}&0&3\end{array}\right]}
=dP_9\x _{P^1} dP_9$.
This Calabi-Yau space
can be constructed  as a $T^6/{\bf Z}_k\x {\bf Z}_{k^{\prime}}$
orbifold, where
we have identified $z_{45}=x_4+ix_5, \; z_{67}=x_6+ix_7, \; z_{89}=x_8+ix_9$.
Locally around the singularities we therefore consider the non-compact
space ${\bf C}^3/{\bf Z}_k\x {\bf Z}_{k^{\prime}}$, where
the ${\bf Z}_k\x {\bf Z}_{k^{\prime}}$ orbifold action, giving a genuine
$\Gamma\subset SU(3)$, is fibered together by corresponding actions giving
an $A_{k-1}$ resp. $A_{k^{\prime}-1}$ singularity in 6789 resp. 4589.
Concretely the corresponding total metric has the following explicit form:
\begin{eqnarray}
ds^2&=& ds_{0123}^2+{1\over {\rm Im}S_{k'}(z_{89})}(d\theta_4-S_{k'}
(z_{89})d\phi_5)
(d\theta_4-\bar S_{k'}(\bar z_{89})d\phi_5)\nonumber\\
&+& {1\over {\rm Im}S_k(z_{89})}
(d\theta_6-S_k(z_{89})d\phi_7)(d\theta_6-\bar S_k(\bar z_{89})d\phi_7)
\nonumber\\
&+&
{\rm Im}S_k(z_{89}){\rm Im}S_{k'}(z_{89})dz_{89}d\bar z_{89}.
\end{eqnarray}
Here $S_k(z_{89})={k\over 2\pi i}\log z_{89}$ and analogously for $S_{k'}$.

This metric descibes a fibration of a $\Gamma$-singularity
in the $z_{4567}$ respectively $z_{67}$ directions over a singular point in
the common base space with coordiates $z_{89}$, just like the $CY^{19,19}$
is a $T^2\times T^2$ fibration over the common base $P^1$.
The $A_{k-1}$ singular point of one $dP_9$ 
direction times the $S^1$ of the remaining $z$-plane
gives the 
complex curve of singularities $z_{67}=z_{89}=0$ resp. $z_{45}=z_{89}=0$ 
intersecting
the $S^5$ relevant to the 
$AdS_5\x S^5/({\bf Z}_{k}\times {\bf Z}_{k'})$ in an
$S^1$ of singularities given by the unit circle in $z_{45}$ resp.
$z_{67}$  (cf. [\ref{HU}]; if $k$ and $k'$ are not coprime there exists
a third curve of singularities).

\subsection{Intersection of D 4-brane with NS 5-branes}

Finally we want 
to describe the metric for the configuration of n D4-branes
between k parallel NS 5-branes in type IIA, 
corresponding to the $N=2$ supersymmetric
$SU(n)^{k-1}$ gauge theory [\ref{Wit}].  
Let us start two steps earlier with D5-branes in type IIB, from
which we get the NS 5-branes by type IIB S-duality, and D3-branes,
being invariant  under the S-duality, from which we get the 
D4-branes in type IIA by T-duality.

Starting with
the intersection of  D5-branes
with  D3-branes we obtain after a S-duality transformation the following
metric which describes the intersection of  NS 5-branes with   D3-branes
 ($e^{2\phi}=H_5$):
\beqa
{\rm NS}5\perp {\rm D}3: \ \ ds^2&=&\frac{1}{\sqrt{H_3}} 
ds^2_{012}+\sqrt{H_3}s^2_{345}+\frac{H_5}{\sqrt{H_3}}ds^2_{6}+
H_5\sqrt{H_3}ds_{789}^2
\eeqa
Next let us perform a T-duality transformation with respect to the $x_3$ 
direction. Under this duality transformation the 
type IIB D3-brane turns into a type IIA D4-brane, i.e. one of the 
transverse 
directions of the D3-brane becomes a worldvolume direction of the 
D4-brane. 
Here we have to assume that $H_3$ is independent of $x_3$ and following 
[\ref{BBJ}] we may write $\tilde{H}_3=H_4$. 
Finally let us apply a T-duality with respect to $x_6$ which leads
to a multi Taub-NUT configuration  
and a fractional D3-brane (where $\omega_i=B_{6i}$ for $i=7,8,9$ and
$e^{2\phi}=1$)
\beqa
ds^2&=&\frac{1}{\sqrt{H_3}}ds_{0123}^2
+\sqrt{H_3}\lbrack ds_{45}^2+
\frac{1}{H_5}(dx_6^2+\vec{\omega}d\vec{x}_{789})^2+
H_5ds^2_{789} \rbrack.
\eeqa
Inspection of this metric clearly exhibits the D3-brane with world volume
along the (0123)-directions as well as the Taub-NUT space in the transversal
directions (6789). 
Now one can proceed as before and `delete' the constant in the harmonic function
$H_5$ via a U-duality transformation. In this way the D3-brane is
localized at 
the $A_{k-1}$ 
singularity in the transversal space (6789).
In addition one can also consider the limit where the constant part
in the harmonic function $H_3$ can be neglected. In this case the 
geometry becomes equivalent to $AdS_5\times S^5/{\bf Z}_{k}$. This space
describes  $N=2$ supersymmetric gauge theories in the large N-limit,
where the theories are supposed to be superconformal.
Finally let us remark that we can also consider the combined system of
D-branes which are positioned in the brane boxes of intersecting
NS 5-branes, as described in section (4.3). The dual geometry of this
set up is then given by D3-branes plus a $\Gamma$ singularity,
which extends into the full transversal space with directions (456789).
At the horizon of the D3-branes the large N-limit of $N=1$
supersymmetric gauge theories, based on the space 
$AdS_5\times S^5/({\bf Z}_{k}
\times {\bf Z}_{k'})$, is obtained.

\section*{Acknowledgements}

We thank K. Behrndt, K. Sfetsos and A. Uranga for discussion. 

\section*{References}
\begin{enumerate}

\item
\label{fivebr}
A. Strominger, Nucl. Phys. {\bf B343} (1990) 167; \\
C.G. Callan, J.A. Harvey and A. Strominger, Nucl. Phys. 
{\bf B359} (1991) 611, Nucl. Phys. {\bf B367} (1991) 60.

\item
\label{rey}
S. Rey, {\em Axionic String instantons and their low energy implications},
published in proceedings to Tuscaloosa Workshop 1989;
{\em On string theory and axionic strings and instantons},
published in DPF Conf. 1991.

\item
\label{schwarz}
J.H. Schwarz, Phys. Lett. {\bf B360} (1995) 13, hep-th/9508143. 

\item
\label{hw}
A. Hanany and E. Witten, {\em Type IIB superstrings, BPS monopoles
and three-dimensional gauge dynamics}, Nucl. Phys. {\bf B492} (1997) 152,
hep-th/9611230.

\item
\label{OV}
H. Ooguri and C. Vafa, {\em Two-dimensional black hole and singularities of
CY manifolds}, Nucl. Phys. {\bf B463} (1996) 55, hep-th/9511164.

\item
\label{frac}
M.R. Douglas and G. Moore, {\em D-branes, quivers and ALE instantons},
hep-th/9603167.

\item
\label{boxes}
A. Hanany and A. Zaffaroni, {\em On the realization of chiral four-dimensional 
gauge theories using branes}, J. High Energy Phys. {\bf 5} (1998) 1,
 hep-th/9801134.

\item
\label{KLS}
A. Karch, D. L\"ust and D. J. Smith, 
{\em Equivalence of Geometric Engineering and Hanany-Witten via 
Fractional Branes}, hep-th/9803232.

\item
\label{HU}
A. Hanany and A. Uranga, {\em Brane Boxes and Branes on Singularities},
hep-th/9805139.

\item
\label{Boon}
H. J. Boonstra, B. Peeters and K. Skenderis, {\em Duality and asymptotic 
geometries}, Phys. Lett. {\bf B411} (1997) 59, hep-th/9706192.

\item
\label{SS}
K. Sfetsos and K. Skenderis, {\em Microscopic derivation of the 
Bekenstein-Hawking entropy formula for non-extremal black holes},
Nucl. Phys. {\bf B 517} (1998) 179, hep-th/9711138.

\item
\label{M}
J. Maldacena, {\em The Large N Limit of Superconformal field theories 
and supergravity}, hep-th/9711200.

\item
\label{BHTZ}
M. Banados, M. Henneaux, C. Teitelboim and J. Zanelli,
{\em Geometry of the $2+1$ Black Hole}
Phys. Rev. {\bf D 48} (1993) 1506, gr-qc/9302012.

\item
\label{sdual}
A. Font, L. Ibanez, D. L\"ust and F. Quevedo, Phys. Lett {\bf B249} (1990) 35;
\\
S. Rey, Phys. Rev. {\bf D43} (1991) 526;\\
A. Sen, Phys. Lett. {\bf B303} (1993) 22;\\
J. Schwarz and A. Sen, Nucl. Phys. {\bf B411} (1994) 35.

\item
\label{CGHS}
C. Callan, S. Giddings, J. Harvey and A. Strominger, 
{\em Evanescent Black Holes}, Phys. Rev. {\bf D 45} (1992) 1005, 
hep-th/9111056.

\item
\label{B}
I. Bakas, {\em Space Time Interpretation of S-Duality and Supersymmetry 
Violations of T-Duality}, Phys. Lett. B343 (1995) 103, hep-th/9410104.

\item
\label{BB}
E. Bergshoeff and K. Behrndt, {\em D-Instantons and asymptotic 
geometries}, hep-th/9803090.

\item
\label{camo}
G. Lopes Cardoso and T. Mohaupt, {\em Dual heterotic black-holes
in four and two dimensions}, hep-th/9806036.

\item
\label{GSVY}
B. Greene, A. Shapere, C. Vafa and S.T. Yau, Nucl. Phys {\bf B 337} (1990)
1.

\item
\label{V}
C. Vafa, {\em Evidence for F-Theory}, Nucl. Phys. {\bf B 469} (1996) 403,
hep-th/9602022.

\item
\label{MS}
J.M. Maldacena and A. Strominger, 
{\em Semiclassical decay of near extremal fivebranes}, hep-th/9710014.

\item
\label{S}
N. Seiberg, hep-th/9608111; hep-th/9609161; hep-th/9705221.

\item
\label{wbh}
E. Witten, Phys. Rev {44} (1991) 314.

\item
\label{HW}
G.T. Horowitz and D.L. Welch, 
{\em Exact Three Dimensional Black Holes in String Theory},
Phys. Rev. Lett. {\bf 71} (1993) 328, hep-th/9302126.

\item
\label{GHM}
R. Gregory, J.A. Harvey and G. Moore, hep-th/9708086.

\item
\label{Sen}
A. Sen, hep-th/9707042; hep-th/9707123.

\item
\label{bianchi}
M. Bianchi, F. Fucito, G.C. Rossi and M. Martinelli,
{\em ALE Instantons in string effective theory},
Nucl. Phys. {\bf  B 440} (1995) 129, hep-th/9409037.

\item
\label{T}
A.A. Tseytlin, {\em Type IIB instanton as a wave in twelve dimensions},
Phys. Rev. Lett. {\bf 78} (1997) 1864, hep-th/9612164.

\item
\label{DS}
D.-E. Diaconescu and N. Seiberg, hep-th/9707158.

\item
\label{A}
P.S. Aspinwall, Phys. Lett. {\bf B 357} (1995) 329.

\item
\label{KKL}
E. Kiritsis, C. Kounnas and D. L\"ust, Journ. of Mod. Phys. {\bf A9}
(1994) 1361, hep-th/9308124 and hep-th/9312143; \\
C. Kounnas, hep-th/9402080.

\item
\label{roc}
S. Gates, C. Hull and M. Rocek, Nucl. Phys. {\bf B258} (1984) 157.

\item
\label{BBJ}
K. Behrndt, E. Bergshoeff and B. Janssen, {\em Intersecting D-branes 
in Ten and Six Dimensions}, hep-th/9604168.

\item
\label{Wit}
E. Witten {\em Solutions Of Four-Dimensional Field Theories Via M Theory},
Nucl. Phys. {\bf B 500} (1997) 3, hep-th/9703166.

\end{enumerate}
\end{document}